\documentclass{article}
\usepackage{spconf,amsmath,amssymb,graphicx}
\usepackage{booktabs}
\usepackage{cite}

\def\titan{TITAN}

\title{\titan{}: Bringing The Deep Image Prior to Implicit Representations}
\name{%
\begin{minipage}{\textwidth}
\centering
Lorenzo Luzi$^1$,
Daniel LeJeune$^1$,
Ali Siahkoohi$^1$,
Sina Alemohammad$^1$,
Vishwanath Saragadam$^1$,\\
Hossein Babaei$^1$,
Naiming Liu$^1$,
Zichao Wang$^{2}$,
Richard G.\ Baraniuk$^1$
\end{minipage}\thanks{This work was supported by NSF grants CCF-1911094, IIS-1838177, and IIS-1730574; ONR grants
N00014-18-12571, N00014-20-1-2534, and MURI N00014-20-1-2787; AFOSR grant FA9550-22-
1-0060; and a Vannevar Bush Faculty Fellowship, ONR grant N00014-18-1-2047. Work done while ZW was at Rice University.}}
\address{$^1$Rice University $\quad\quad\quad$ $^2$Adobe Research\\
}

\usepackage{hyperref}
\usepackage[noabbrev, capitalise]{cleveref}

\usepackage{siunitx}
\usepackage{tikz}
\usepackage{circuitikz}
\usepackage{pgfplots}
\usetikzlibrary{positioning, fit, shapes.geometric}
\pgfplotsset{compat=1.15}

\usepackage{amsmath,amsfonts,bm, bbm}

\def\eqref#1{equation~\ref{#1}}

\def\1{\bm{1}}

\def\vv{{\bm{v}}}

\def\vx{{\bm{x}}}

\def\mB{{\bm{B}}}
\def\mC{{\bm{C}}}

\def\mI{{\bm{I}}}

\def\mR{{\bm{R}}}
\def\mS{{\bm{S}}}

\def\mU{{\bm{U}}}

\def\mW{{\bm{W}}}

\def\mY{{\bm{Y}}}

\DeclareMathAlphabet{\mathsfit}{\encodingdefault}{\sfdefault}{m}{sl}
\SetMathAlphabet{\mathsfit}{bold}{\encodingdefault}{\sfdefault}{bx}{n}

\def\sR{{\mathbb{R}}}

\usepackage{mdframed}

\usepackage{accents}
\newlength{\dhatheight}

\begin{document}

\maketitle
\begin{abstract}

We study the interpolation capabilities of implicit neural representations (INRs) of images. In principle, INRs promise a number of advantages, such as continuous derivatives and arbitrary sampling, being freed from the restrictions of a raster grid. However, empirically, INRs have been observed to poorly interpolate between the pixels of the fit image; in other words, they do not inherently possess a suitable prior for natural images. In this paper, we propose to address and improve INRs' interpolation capabilities by explicitly integrating image prior information into the INR architecture via deep decoder, a specific implementation of the deep image prior (DIP). Our method, which we call TITAN, leverages a residual connection from the input which enables integrating the principles of the grid-based DIP into the grid-free INR. Through super-resolution and computed tomography experiments, we demonstrate that our method significantly improves upon classic INRs, thanks to the induced natural image bias. 
We also find that by constraining the weights to be sparse, image quality and sharpness are enhanced, increasing the Lipschitz constant.

\end{abstract}

\section{Introduction}
\label{sec:intro}

Implicit neural representations (INRs) aim to represent images with a differentiable (neural network) function instead of the traditional discrete raster image of pixel point values in a 2-dimensional grid. Concretely, an INR might learn a function mapping from an arbitrary location in 2D, represented by $(x,y)$, to the $(r,g,b)$ values in the image:
\begin{align}
    \mI: \sR^2 \rightarrow \sR^3, \quad \mI(x,y) = (r,g,b)\,.
\end{align}
Thanks to their desirable characteristics including being grid-free, continuous, and differentiable, INRs have been an increasingly popular and integral component for a wide range of computer graphics and vision applications such as super-resolution~\cite{2020arXiv201209161C,2021arXiv210312716X}, 3D scene rendering~\cite{chen2018implicit_decoder,Occupancy_Networks,Park_2019_CVPR,mildenhall2020nerf}, and image generation~\cite{inr_gan,RottShaham2020ASAP,2021arXiv210204776D,anokhin2020image}. Among the most famous INRs is SIREN~\cite{sitzmann2020implicit}, which consists of a feed-forward neural network with sinusoidal activation functions.
Others have used wavelet activations instead of sinusoidal ones with significant success~\cite{saragadam2023wire}.

The most common class of INR models only use a single datum and do not require a training data corpus. Models trained with a collection of images may suffer from generalization problems due to under-specialization. Interpolating a single image without training makes it more suitable and safer in high-stakes applications such as computed tomography (CT) imaging, where images may vary significantly from patient to patient due to their different anatomical structures~\cite{gong2018learning}. Moreover, data-free INRs can be deployed to challenging imaging tasks in data-starved environments.

\begin{figure}[t]
    \centering
    \def\figWidth{.2}
    
    \newcommand{\two}[2]{
    {\small
    \begin{tabular}{@{}c@{}} %
    #1 \\ #2
    \end{tabular}
    }}
    
    \begin{tikzpicture}[node distance=0.2cm and 0.2cm,
    every node/.style={inner sep=0cm,outer sep=0.02cm}]
    \node (top1) [] {\two{\\Ground Truth}{\includegraphics[width=\figWidth\textwidth]{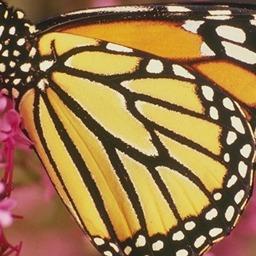}}};
    
    \node (top2) [right= of top1] {\two{\\Downsampled $4\times$}{\includegraphics[width=\figWidth\textwidth]{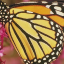}}};
    
    \node (bot1) [below= of top1] {\two{SIREN\\$19.6$ dB PSNR, 0.75 SSIM}{\includegraphics[width=\figWidth\textwidth]{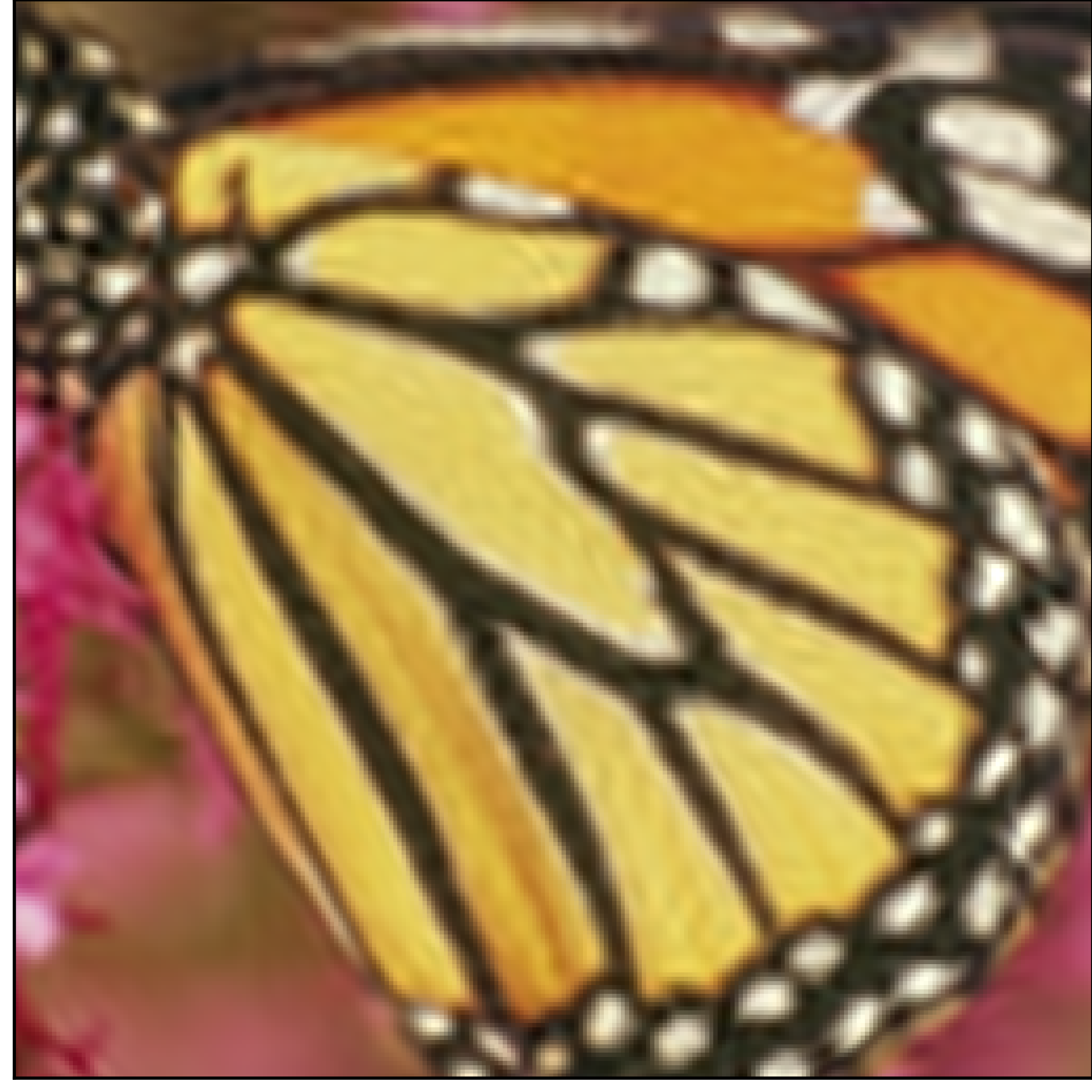}}};
    
    \node (bot2) [right= of bot1] {\two{\titan{}\\$21.7$ dB PSNR, 0.83 SSIM}{\includegraphics[width=\figWidth\textwidth]{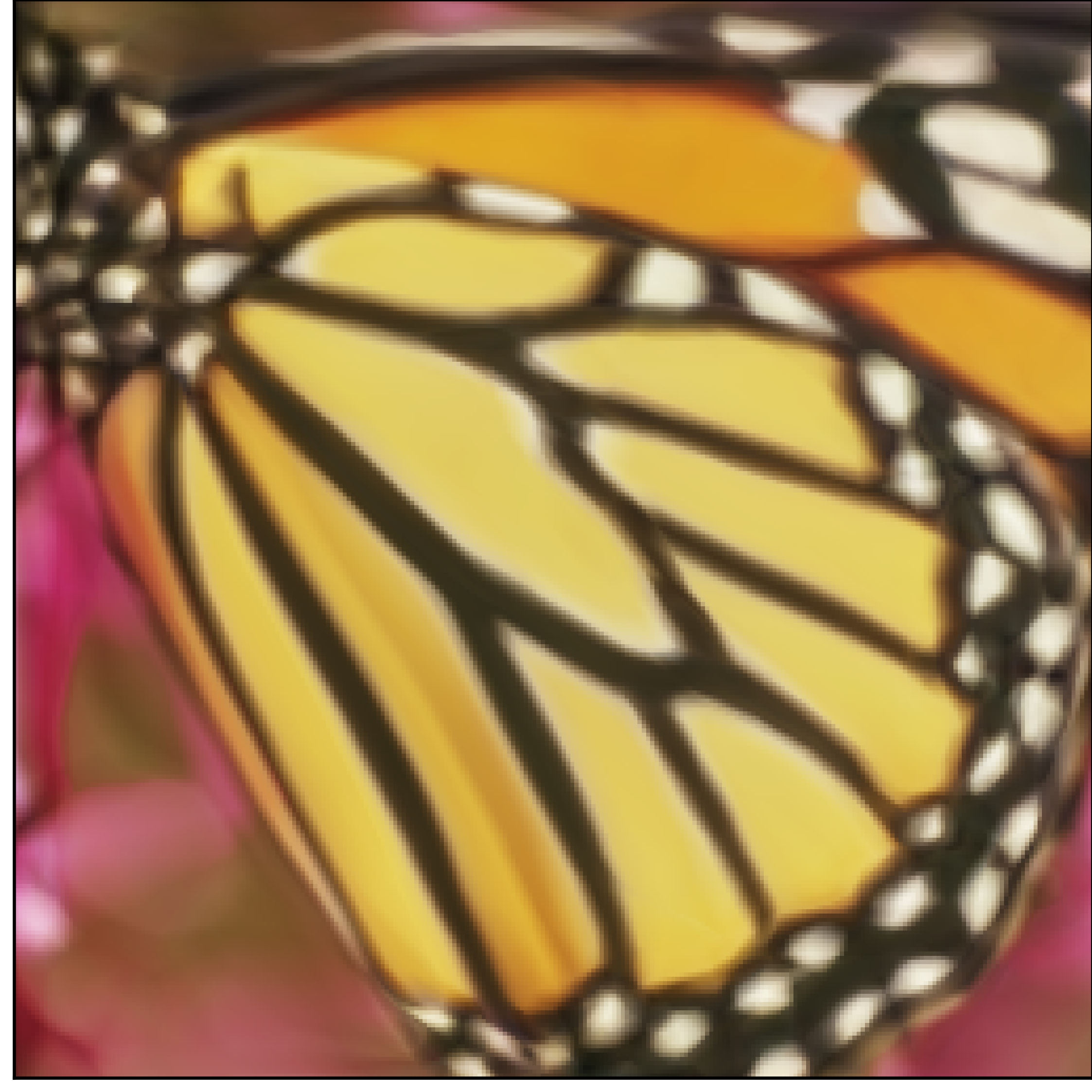}}};
    
    \end{tikzpicture}
    
    \caption{Using \titan{} we outperform SIREN on a $4\times$ super-resolution task. \titan{} here has approximately the same parameters ($\approx 112\,\mathrm{k}$) as SIREN ($\approx 133\,\mathrm{k}$) due to the weights being sparse.}
    \label{fig:exp superresolution}
\end{figure}

However, a notable drawback of INRs is that they are often poor interpolators. Prior work~\cite{sitzmann2020implicit, Yuce_2022_CVPR} has empirically demonstrated that INRs often fail to represent images in finer scales; see \cref{fig:exp superresolution} for an illustration of the unsatisfactory image representation performance of SIREN in a super-resolution case study. This practical deficiency is concerning because the continuous, grid-free nature of INRs should enable image representation at arbitrary scale and resolution.

In this paper, we consider INRs' grid-free interpolation capabilities without relying on other (neural) image models which use discrete pixel representations. Instead, we take inspiration from the deep decoder~\cite{heckel2019decoder} and design an INR architecture which is both grid-free and leverages the deep image prior~\cite{lempitsky2018prior}.

\subsection{Contributions}
We propose a new method to significantly improve INR interpolation capabilities. Our method is inspired by
the deep image prior (DIP)~\cite{lempitsky2018prior}, an observation that certain architectural choices in generative networks bias them towards natural images. We postulate that
INRs suffer from poor interpolation capabilities because they lack such architectural inductive biases.

To this end, we integrate DIP information into an INR using deep decoder, a simplified implementation of DIP, which automatically enforces strong prior information for a given image with a powerful but simple neural network. 
We add residual connections from the input that enable us to integrate the principles of deep decoder, which normally operates in the pixel representation of images, into an INR, which operates in the grid-free functional representation of images. This yields an image representation with powerful inductive biases.
Through a series of experiments on image super-resolution and CT recovery, we show that our method, TITAN, outperforms the 
INR methods without image prior by a large margin, 
demonstrating the importance and the promise in leveraging image prior information in INRs to turn them into powerful interpolators.

\section{Background}
\label{sec:background}

Our method fuses DIP techniques, specifically deep decoder, with INRs that allow grid-free image inference.

\subsection{Deep image priors and deep decoder}
\label{sec:dip}
Deep image prior~\cite{lempitsky2018prior} proposes an untrained network to capture image statistics prior for inverse problems, such as denoising, image inpainting, and super-resolution. DIP takes a random vector as input and outputs the image prior using a UNet-like~\cite{unet} network such as hourglass network or encoder--decoder with skip connections. It surprisingly shows that the simple structure of a deep convolution network is sufficient to enforce reasonable image priors without the need for extensive training on large datasets. Inspired by the work of DIP, we propose our method \titan{} to learn a high-quality implicit representation from a single image.

\begin{figure*}[t]
    \centering
    \includegraphics[width=0.9\linewidth]{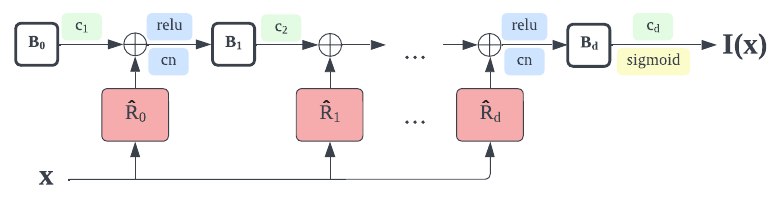}
    \vspace{-5mm}
    \caption{A diagram of the \titan{} architecture. The input coordinate $\vx$ passes through the residual blocks $\widehat \mR_i$ as a substitute for upsampling but otherwise the output pixel value follows the deep decoder architecture.}
    \label{fig:titan}
\end{figure*}

\noindent\textbf{Deep decoder.}
The deep decoder~\cite{heckel2019decoder} is a raster-based (i.e., defined on a grid) implicit image representation that distills the ideas behind the DIP to its essential components. It differs from the DIP in that it has no skip connections and no convolutional operations, but rather limits inter-pixel interactions to only upsampling operations.
Starting with a fixed $n_0 \times n_0 \times k_0$ random
``image'' $\mB_0$, the architecture layers proceed via the following recursive relation:
\begin{align}
    \label{eq:deep-decoder-layer}
    \mB_{i+1} = \mathrm{cn}(\mathrm{relu}(\mU_i \mB_i \mC_i)).
\end{align}
The nonlinearity is $\mathrm{relu}(t) = \max\{0, t\}$, and $\mathrm{cn}$ is a channel-wise normalization, also called a one-dimensional batch normalization~\cite{bn}, with optional learned parameters. 
For $\mB_{i + 1}$ of shape $n_i \times n_i \times k_i$,  $\mU_i$ is a fixed bilinear upsampling operator that lifts each channel from $n_i \times n_i$ to $n_{i + 1} \times n_{i + 1}$, where $n_{i + 1} = 2 n_i$.
The weights $\mC_i \in \sR^{k_i \times k_{i +1}}$ are learned parameters which mix the $k_i$ channels to $k_{i + 1}$ channels. 
The final result is an $n_d \times n_d \times k_d$ raster image $\mI$ formed by
\begin{align}
    \mI = \mathrm{sigmoid}(\mB_d \mC_d),
\end{align}
where $\mathrm{sigmoid}(t) = 1 / (1 + e^{-t})$. Training the weights $\mC_i$ via gradient descent results in a robust image prior.

\section{Method}
\label{sec:method}

We would like to take advantage of the inductive biases of the deep image prior and deep decoder in an INR, but the convolutional and upsampling operations that incorporate inter-pixel interactions preclude a direct implementation. 
To address this, we propose \titan{}, which stands for Deep Implicit Decoder Network,\footnote{We prefer the nearly phonetically identical TITAN to DIDN because it is both beautiful and good.} 
which implements a deep decoder architecture as an INR via a careful replacement of the upsampling operator with spatial residual connections. A diagram of \titan{} is shown in Figure~\ref{fig:titan}.

We seek an implicit representation $\mI_\theta \colon \sR^2 \to \sR^{k_d}$ with parameters $\theta \in \sR^p$ of an image such that given a coordinate $\vx \in \sR^2$, $\mI_\theta(\vx)$ returns the pixel values at that coordinate across the $k_d$ channels. We start the same as the deep decoder with $n_0 = 1$: we let $\mB_0 \in \sR^{k_0}$ be a vector representing a constant image of $k_0$ channels. We wish to implement the next layer of the deep decoder, which according to \eqref{eq:deep-decoder-layer} should look something like %
\begin{align}
    \mB_{i+1}(\vx) = \mathrm{cn}(\mathrm{relu}(\mC_i [(\mU_i \mB_i)(\vx)])).
\end{align}
However, in order to have a one-to-one implementation of deep decoder as an INR, we would need $(\mU_i \mB_i)(\vx)$ to be the upsampled version of the image $\mB_i \colon \sR^2 \to \sR^{k_i}$ evaluated at $\vx$. Upsampling does not really have meaning for non-raster images; for raster images, however, it is the inverse operation of downsampling, which is analogous to blurring. Thus, $\mB_i(\vx)$ should be similar to $(\mU_i \mB_i)(\vx)$, but with less detail. We can therefore define the upsampling \emph{residual} 
\begin{align}
    \mR_i(\vx) \triangleq (\mU_i \mB_i)(\vx) - \mB_i(\vx).
\end{align}
What we do in \titan{} is explicitly approximate $\mR_i(\vx)$ via a simple nonlinear function, like a small SIREN network:
\begin{align}
    \widehat{\mR}_i(\vx) = g(\alpha_i(\mW_i \vx + \vv_i)) / d,
\end{align}
where $g$ is an element-wise nonlinearity that captures spatial frequency information---we use $g(t) = \sin(t)$; $\alpha_i$ is a fixed frequency scaling parameter; and dividing by $d$ ensures that the total contribution of the residuals is fixed. We then have the TITAN layer update function
\begin{align}
    \mB_{i+1}(\vx) = \mathrm{cn}(\mathrm{relu}(\mC_i \mB_i(\vx) + \widehat{\mR}_i(\vx))),
\end{align}
with final output
\begin{align}
    \mI_{\theta}(\vx) = \mathrm{sigmoid}(\mC_d \mB_d(\vx)).  
\end{align}
In our experiments, we are not concerned with ensuring the differentiability of our TITANs, so we use the $\mathrm{relu}$ nonlinearity even though it is non-differentiable. If differentiability is necessary, the nonlinearity can be replaced by a differentiable alternative such as the softplus function.

\section{Experiments}
\label{sec:exp}

For all experiments, we set the frequency scaling parameters of TITAN to $\alpha_i = 4(i + 1)$ so that higher frequencies are captured in deeper layers, and we let $k_0 = k_1 = \ldots k_{d-1} = 100$.
Weight initializations for TITAN are the default PyTorch initialization. We use the Adam~\cite{kingma2014adam} optimizer with learning rate $10^{-3}$ for SIREN and TITAN and learning rate $10^{-2}$ for DIP unless otherwise specified. Code can be found at \url{https://github.com/dlej/titan-implicit-prior}

\subsection{TITAN for super-resolution}

We perform $4 \times$ super-resolution of a $256 \times 256$ image downsampled to $64 \times 64$ and show the result in \cref{fig:exp superresolution}. For TITAN, we use depth $d = 10$, and for SIREN we use $d = 2$. We use the AdaBreg \cite{JMLR:v23:21-0545} optimizer for sparsity.

As we can see, TITAN results in a much sharper image than SIREN, which suffers from Gibbs-like ringing artifacts. For this task, DIP achieves a PSNR of 23.9 and 0.89 SSIM, so TITAN is a solid step in the direction of  inductive image biases for INRs.

\begin{figure}[t]
    \centering
    \def\figWidth{.2}
    
    \newcommand{\two}[2]{
    {\small
    \begin{tabular}{@{}c@{}} %
    #1 \\ #2
    \end{tabular}
    }}
    
    \begin{tikzpicture}[node distance=0.3cm and 0.2cm,
    every node/.style={inner sep=0cm,outer sep=0.02cm}]
    \node (top1) [] {\two{\\Ground Truth}{\includegraphics[width=\figWidth\textwidth]{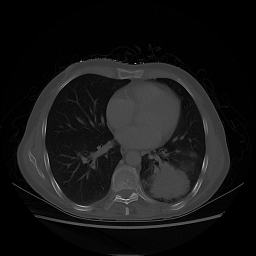}}};
    \node (top2) [right= of top1] {\two{DIP\\$30.5$ dB PSNR, 0.92 SSIM}{\includegraphics[width=\figWidth\textwidth]{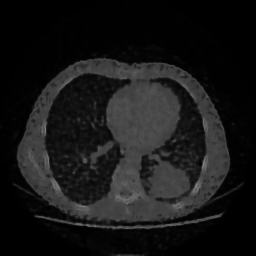}}};
    \node (bot1) [below= of top1] {\two{SIREN\\$28.5$ dB PSNR, 0.81 SSIM}{\includegraphics[width=\figWidth\textwidth]{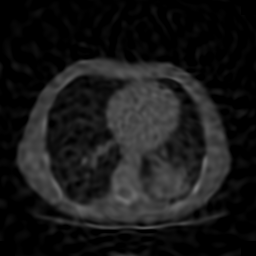}}};
    \node (bot2) [right= of bot1] {\two{TITAN\\$29.9$ dB PSNR, 0.89 SSIM}{\includegraphics[width=\figWidth\textwidth]{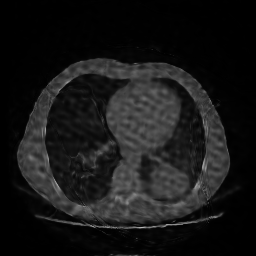}}};
    \end{tikzpicture}

    \caption{For number of measurements $m = 30$, \titan{} outperforms SIREN by leveraging the deep image prior's inductive bias. Thus, \titan{} has all the benefits of being an implicit representation network with only a small ($\approx 0.6$dB) cost to performance. Also \titan{} begin to outperform SIREN and DIP for larger number of measurements. }
    \label{fig:exp CT}
\end{figure}

\subsection{TITAN for computed tomography}
For the ground truth CT image $\mY \in \mathbb{R}^{W\times H}$ in \cref{fig:exp CT}, we take noisy measurements of the following form:
\begin{align}
    \mB = \text{RT}(\mY,m) + \mS \in \mathbb{R}^{m \times H}
\end{align}
Where $\text{RT}(\mY,m) \colon \mathbb{R}^{W \times H} \to \mathbb{R}^{m\times H}$ is the Radon Transform  with $m$ uniformly spaced samples from $0$ to $\pi$ and $\mS_{ij} \overset{\mathrm{i.i.d.}}{\sim} \mathcal{N}(0, \sigma^2) $ is random Gaussian noise with standard deviation $\sigma = 2$. We optimize the following cost function
\begin{align}
    \mathcal{L}(\theta) = \|  \text{RT}(\mI_{\theta}(\vx), m) - \mB  \|_F^2
    \label{eq:opt}
\end{align}
using Adam with cosine annealing rate.
Results of all methods for several numbers of measurements $m$ are shown in Table~\ref{tab:ct table}. TITAN provides a significantly better implicit representation for solving this inverse problem than SIREN, nearly matching or even surpassing the DIP at all measurement levels.

\begin{table}[h]
    \aboverulesep=0.1ex
    \belowrulesep=0.1ex

    \centering
    \begin{tabular}{cc|cccc}
        \toprule
        Method & Metric & $m=30$ & $40$ & $50$ & $100$ \\
        \midrule
        SIREN & PSNR & 28.5 & 29.9 & 31.3 & 33.2 \\
        & SSIM & 0.81 & 0.86 & 0.90 & 0.95 \\
        \midrule
        TITAN & PSNR & 29.9 & 31.5 & \textbf{32.5} & \textbf{36.1} \\
        & SSIM & 0.89 & 0.91 & 0.94 & 0.97 \\
        \midrule
        DIP & PSNR & \textbf{30.5} & \textbf{31.8} & 31.9 & 32.6 \\
        & SSIM & 0.92 & 0.94 & 0.95 & 0.96
        \\ \bottomrule
    \end{tabular}
    \caption{Results showing that TITAN outperforms SIREN on computed tomography tasks with varying number of measurements $m$.}
    \label{tab:ct table}
\end{table}

\subsection{The effect of sparsity}

Partly motivated by the success of INRs in data compression~\cite{Vishwanath2020}, we propose to compensate for the larger parameter count of \titan{} compared to SIREN via sparsity-promoting INF optimization (c.f.\ equation~\ref{eq:opt}). We achieve this via an optimization approach~\cite{JMLR:v23:21-0545} based on linearized Bregman iterations~\cite{Yin2008}. Unlike pruning methods~\cite{NIPS1989_6c9882bb}, this Bregman learning method initializes the INR with a few nonzero weights, successively adding limited nonzero weights throughout optimization. We tune the final sparsity factor of the weights via a hyperparameter, which controls the initialization sparsity factor. Our empirical results for image super-resolution and computed tomography demonstrate both quantitative, measured via PSNR, and perceptual improvement of the resulting images, specifically complementing the inductive bias of \titan{} towards attenuating of Gibbs ringing artifacts typically observed when using SIREN. Surprisingly, the Bregman learning algorithm was not able to sparsify the weights of SIREN when initialized with the same sparsity factor as \titan{}.

\subsection{The effect of sparsity on Lipschitz constants}

We investigate whether the smoothness induced by the sparsity yields an implicit model which has a lower Lipschitz constant than the non-sparse model. To do this, we focus on the super-resolution problem and generate a fine grid of $256\times 256$ pixel locations and their corresponding model outputs. We follow this by calculating the largest singular value of the Jacobian of our implicit model at each pixel, computed via backpropagation. Finally, we take the largest of these values to obtain the Lipschitz constant of our INR.

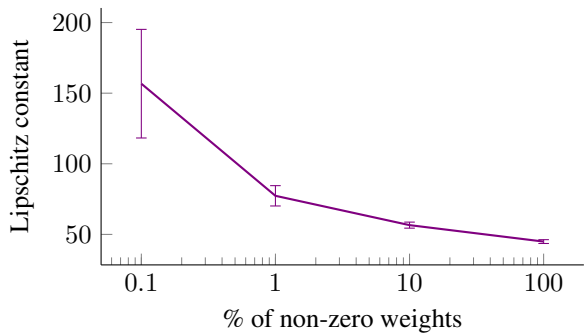
\begin{figure}[!ht]
    \centering
    \pgfplotsset{
        axisStyleLine/.style={
          width=8cm,
          height=5cm,
          axis x line*=bottom,
          axis y line*=left,
          }
        }
        \tikzstyle{plotStyleLineYerr}=[
                    color=violet, 
                    mark={},
                    thick]
        \tikzstyle{plotStyleLine}=[
                    color=purple, 
                    mark={},
                    thick]
        \tikzstyle{plotStylePoint}=[
                    color=Red, 
                    only marks,
                    mark=o,
                    ultra thick,
                    mark size=2pt
                    ]
    \begin{tikzpicture}
    \pgfplotsset{
  log ticks with fixed point,
}
        \begin{semilogxaxis}[
            axisStyleLine,
            ylabel={Lipschitz constant},
            xlabel={\% of non-zero weights}]
            \addplot+[
                plotStyleLineYerr,
                error bars/.cd,
                y explicit,
                y dir=both] 
            table [
                x index=0, 
                y index=1,
                y error index=2,
                col sep=comma] {csv/lipschitz.csv};
        \end{semilogxaxis}
    \end{tikzpicture}
    \caption{As the number of non-zero weights of \titan{} at initialization increases, the Lipschitz constant decreases; hence more sparse solutions tend to have larger Lipschitz constants since they are sharper. The sparsity of \titan{} after training does not change much for a given initialization, and so we group them together and plot the average Lipschitz constant with its standard error for $n=10$ random seeds. }
    \label{fig:lipschitz}
\end{figure}

Our findings are somewhat counter-intuitive: the Lipschitz constant decreases as we increase the number of non-zero weights at the beginning of training as shown in \cref{fig:lipschitz}. This is especially counter-intuitive because lower Lipschitz constants are correlated with better generalization performance~\cite{ramasinghe2021beyond}, and we see the opposite here. However, this happens because the largely smooth \titan{} representation we see has sharp edges, which induces a large Lipschitz constant.

\section{Conclusion}
\label{sec:conclusion}

We have demonstrated that it is possible to incorporate the inductive biases of the DIP into an INR, which we have implemented via residual connections in the place of the upsampling operator of a deep decoder. Complemented with sparsity-promoting optimization over weights, our proposed approach mitigates common perceptual artifacts when deploying INRs while maintaining a low parameter count.  INRs with robust inductive biases will enable their deployment to solving hard imaging problems in resource-constrained settings.

\vfill\pagebreak

\bibliographystyle{IEEEbib}
\bibliography{refs}

\end{document}